# Multigrid Methods for Propagators in Lattice Gauge Theories [*]

Thomas Kalkreuter [**]

*Fachbereich Physik (Computational Physics)*
*Humboldt-Universität*
*Invalidenstraße 110*
*D-10099 Berlin, Germany*

**Abstract**

Multigrid methods were invented for the solution of discretized partial differential equations in ordered systems. The slowness of traditional algorithms is overcome by updates on various length scales. In this article we discuss generalizations of multigrid methods for disordered systems, in particular for propagators in lattice gauge theories. A discretized nonabelian gauge theory can be formulated as a system of statistical mechanics where the gauge field degrees of freedom are $SU(N)$ matrices on the links of the lattice. These $SU(N)$ matrices appear as random coefficients in Dirac equations. We aim at finding an efficient method by which one can solve Dirac equations without critical slowing down. If this could be achieved, Monte Carlo simulations of Quantum Chromodynamics (the theory of the strong interaction) would be accelerated considerably. In principle, however, the methods discussed can be used in arbitrary space-time dimension and for arbitrary gauge group. Moreover, there are applications in multigrid Monte Carlo simulations, and for the definition of block spins and blocked gauge fields in Monte Carlo renormalization group studies. As a central results it was found that *the geometric multigrid method works in principle in arbitrarily disordered gauge fields*. Finally, an overview is given of other approaches to the propagator problem in lattice gauge theories.

---

# 1 Introduction

The second part of the title of this contribution indicates a rather specialized topic in elementary particle physics ("propagators in lattice gauge theories"), but I hope the majority of readers from other fields can benefit from the first part ("multigrid methods") which will be discussed from a general point of view. The aim of multigrid (MG) methods is to beat critical slowing down in nearly critical systems, i.e. to maintain fast convergence when long range correlations appear. Actually in this article we should speak of deterministic MG methods, in contrast to stochastic MG methods which will not be covered here.

Let us consider the general problem of solving a very large linear system of equations

$$D\phi = f , \qquad (1)$$

where $D$ is a sparse $N \times N$ matrix, and the r.h.s. $f$ is given. We will also assume that $D$ is hermitean and positive. One can think of (1) as a discretization of a partial differential equation on a lattice. Classical solvers are for instance successive over-relaxation (SOR) or conjugate gradient (CG) algorithms [10]. In cases where $D$ has no low eigenvalue or a small condition number, classical iteration algorithms converge fast. However, in situations of practical interest, the contrary is true. As $N$ gets bigger and bigger in order to obtain a good approximation to the solution of a continuum differential equation, the computational effort per degree of freedom for solving (1) diverges. This phenomenon is called *critical slowing down* (CSD).

MG algorithms were invented to circumvent the problem of CSD. For the solution of discretized partial differential equations in ordered systems, MG algorithms are state-of-the-art solvers. They overcome the slowness of traditional algorithms by updates on various length scales. Introductions to this subject can be found in the classical papers of Brandt, and Stüben and Trottenberg [4], in the textbook by Hackbusch [12], and at a very elementary level in the book by Briggs [6].

This lecture is organized as follows. We discuss in Sec. 2 how MG methods eliminate CSD completely. In Sec. 3 we turn to an MG method which was given the name "ground-state projection MG"; it is based on the existence of a Hamiltonian or action which defines the problem. Finally, in Sec. 4 MG approaches to the propagator problem in elementary particle physics are summarized.

# 2 Principle of Multigrid Methods

Let us start by recalling briefly traditional iterative methods. Classical relaxation algorithms such as SOR [10] generate a sequence $\{\phi^{(n)}\}$, $n = 0, 1, 2, \ldots$, of approximate solutions of Eq. (1). This sequence is obtained through an affine fixed point iteration with some iteration matrix $M$. Every approximation $\phi^{(n)}$ differs from the exact solution by an *error*

$$e^{(n)} \equiv \phi - \phi^{(n)} . \qquad (2)$$

The error gets deamplified by $M$: $e^{(n+1)} = Me^{(n)} = M^{n+1}e^{(0)}$. Asymptotically the error decays exponentially with a *relaxation time* $\tau$, defined by $\tau = -1/\ln \rho(M)$, where $\rho$ denotes the spectral radius of $M$. Numerically $\tau$ is determined by monitoring ratios $\|r^{(n+1)}\|/\|r^{(n)}\|$, which approach $\rho(M)$ asymptotically. $r^{(n)}$ denotes the *residual*,

$$r^{(n)} \equiv f - D\phi^{(n)} \ . \tag{3}$$

Monitoring the exponential decay of the residual leads to the same relaxation time as monitoring of the norm of errors $\|e^{(n+1)}\|/\|e^{(n)}\|$, because the matrix which damps the residual is $DMD^{-1}$, and this matrix is similar to the iteration matrix. For any positive definite operator $D$, SOR converges for arbitrary initial $\phi^{(0)}$ if and only if $0 < \omega < 2$, where $\omega$ is the relaxation parameter [10]. In our case $D$ is positive definite, so we have an algorithm which solves Eq. (1). However, $\tau$ diverges as $D$ becomes critical, i.e. in the interesting case that the lowest eigenvalue of $D$ is close to zero. This CSD of the algorithm can be tackled successfully by MG methods.

The basic observation for MG methods is the following. Classical relaxation algorithms are effective in *smoothing the error*, but as soon as the error is smooth (on the length scale of the given discretization) it is reduced only very slowly because of CSD. However, a smooth function can be represented very well on a coarser lattice (with fewer degrees of freedom). Suppose for instance that the values of a lattice function are given only on every second site. Then, if one knows that the function is smooth, one can reconstruct it to a good accuracy by interpolation. We will now explain these ideas in more detail.

For simplicity we assume that (1) is a discretization on a $d$-dimensional hypercubic lattice $\Lambda$ of lattice spacing $a$. In the MG approach for solving Eq. (1) we divide $\Lambda$ into hypercubes ("blocks") $x$ consisting of $L_b^d$ sites $z \in \Lambda$, with $L_b \in \mathbb{N}$, [1]) typically $L_b = 2, 3$. We identify each such hypercube $x$ with the site $\hat{x}$ at its center, and we write $z \in x$ if $z$ is a site in block $x$. (If $L_b$ is even there is no such distinguished center. Then define arbitrarily an $\hat{x}$ in one block; this defines the other block centers by the requirement that the block lattice is regular.) The sites $\hat{x}$ form the first block lattice $\Lambda^1$ with lattice spacing $L_b a$, and so on. This yields a sequence of lattices $\Lambda = \Lambda^0, \Lambda^1, \Lambda^2, \ldots$ of increasing lattice spacing $a_i$, viz. $a_{i+1} = L_b a_i$ with $a_0 = a$. (One may also use different blocking factors $L_b$ on different layers of the MG).

After some relaxation sweeps on $\Lambda^0$ one gets an approximation $\phi^{(n)}$ to $\phi$ which differs from the exact solution by an error $e^{(n)} \equiv e_0$, Eq. (2). The error satisfies the *residual equation*

$$D_0 e_0 = r_0 \ , \tag{4}$$

where $r_0 \equiv r^{(n)}$, and we wrote $D_0$ for $D$. If $e_0$ is smooth, it is determined to a very good accuracy by a function $e_1$ on the next coarser lattice $\Lambda^1$, and can be represented in the form

$$e_0 = \mathcal{A} e_1 \tag{5}$$

---
[1]) The requirement $L_b \in \mathbb{N}$ is not compulsory in general; see e.g. [11, 17].

with an interpolation map [2] $\mathcal{A}$ which should be so chosen such that it maps functions on $\Lambda^1$ into smooth functions on $\Lambda^0$. Conversely, $e_1$ should be obtainable from $e_0$ with the help of an averaging map $C$. Usually one requires some normalization condition for the kernels, e.g. $CC^* = 1\!\!1$ and $C\mathcal{A} = 1\!\!1$, or $< C, C > = 1$ and $< C, \mathcal{A} > = 1$ with a suitable scalar product. Inserting Eq. (5) into Eq. (4) and acting on the result with $C$, we see that $e_1$ will satisfy the equation

$$D_1 e_1 = r_1 \tag{6a}$$

with

$$D_1 = CD_0\mathcal{A} \ , \quad r_1 = Cr_0 \ . \tag{6b}$$

The choice of the coarse grid operator $D_1$ in (6b) is called "Galerkin choice". In general $D_1$ can also be defined differently, see e.g. [4, 12, 11]. The problem has been reduced to an equation on the coarser lattice $\Lambda^1$ which has fewer points. If there is still too much CSD at this level, one may repeat the procedure, going to coarser and coarser lattices. The procedure stops, because an equation on a "lattice" with only a single point is easy to solve.

After solution of Eq. (6a) one replaces $\phi^{(n)} \mapsto \tilde{\phi}^{(n)} \equiv \phi^{(n)} + \mathcal{A}e_1$. Note that the residual of the corrected approximation $\phi^{(n)} + \mathcal{A}e_1$ vanishes when it is transferred back to $\Lambda^1$. If $\mathcal{A}e_1$ were equal to $e_0$, then $\tilde{\phi}^{(n)}$ would be the solution of Eq. (1). In practice, however, one has to repeat the procedure: do relaxation with $\tilde{\phi}^{(n)}$, solve the residual equation for the new error, etc.

The reason for the efficiency of the MG method is that with a suitable choice of $C$, $\mathcal{A}$, $D_1$ etc. only a few iterations are needed to reduce the error to a small value, independent of the degree of criticality of the problem. In other words, CSD is completely eliminated by MG. This statement has been known to be true in ordered systems, and we will see below what was found for disordered systems. A crucial problem is how to define and exhibit smooth functions in a disordered context, i.e. when translation symmetry is strongly violated.

Another advantage of the MG method is that the computational labor for one MG iteration is comparable to that of conventional relaxation, irrespective of the total number of layers. For details of this work estimate see Refs. [4, 12, 6, 11]. There are further terms which are relevant in MG algorithms for which we have to refer to the literature. These terms include the "cycle control parameter $\gamma$", the notion of "V-cycles" ($\gamma = 1$) and "W-cycles" ($\gamma = 2$), etc.

In order to specify an MG algorithm, we have to make a specific choice for the restriction operator $C$ and for the interpolation operator $\mathcal{A}$. These operators will be defined by their integral kernels $C(x,z)$ and $\mathcal{A}(z,x)$. ($z \in \Lambda^j$, $x \in \Lambda^{j+1}$, e.g. $z \in \Lambda^0$, $x \in \Lambda^1$.) For reasons of practicality one must require that $\mathcal{A}(z,x) = 0$ unless $z$ is near $\hat{x}$.

---

[2] The notation used here follows Mack's Cargèse lectures [25] which were inspired by rigorous works in constructive quantum field theory. In particular our notation for retsriction and interpolation operators differs from the one used in the mathematical literature. When $\mathcal{H}^j$ denotes the space of functions on $\Lambda^j$, then we have

$$\mathcal{A}^j \ : \ \mathcal{H}^j \to \mathcal{H}^{j-1} \quad : \text{interpolation operator}$$
$$C^j \ : \ \mathcal{H}^{j-1} \to \mathcal{H}^j \quad : \text{restriction operator}$$

For notational simplicity we will frequently omit lattice indices $j$.

Adopting a variational point of view, one uses the fact that solving Eq. (1) is equivalent to minimizing the energy functional

$$K[\phi] = \frac{1}{2} < \phi, D\phi > - < \phi, f > . \qquad (7)$$

If one requires that $K$ is lowered as far as possible in every MG correction step $\phi^{(n)} \mapsto \phi^{(n)} + \mathcal{A}e_1$ (under the restriction that $\phi^{(n)}$ is fixed), it follows that the averaging map $C$ and the interpolation map $\mathcal{A}$ are adjoints of one another [4, 12, 11]:

$$C = \mathcal{A}^* . \qquad (8)$$

Then we can define the coarse grid operator simply as $D_1 = CD_0C^* \equiv CDC^*$, and "all" we have to do is to specify $C$. (The integral kernel of the adjoint averaging operator is $C^*(z,x) = C(x,z)^\dagger$, where $\dagger$ denotes the hermitean conjugate of a matrix.)

Finally, let us mention an idealized or "optimal" MG method. Its use as a starting point in numerical work was proposed by Mack [25]. Given the averaging kernel $C$, there exists an ideal choice of the interpolation kernel $\mathcal{A}$. It is determined as follows. For every function ("block spin") $\Phi$ on $\Lambda^1$, $\phi = \mathcal{A}\Phi$ minimizes the action $< \phi, D\phi >$ subject to the constraint $C\phi = \Phi$. With this choice of $\mathcal{A}$, $D_1 = CD_0\mathcal{A}$ is guaranteed to be self-adjoint. A good "choice of block spin", i. e. of $C$, is characterized by the fact that the ideal kernel $\mathcal{A}(z,x)$ associated with it has good locality properties. This means that $\mathcal{A}(z,x)$ is big for $z \in x$, and decays exponentially in $|z - \hat{x}|$ with decay length one block lattice spacing.

The above characterization of $\mathcal{A}$ is equivalent to saying that with the ideal choice of $\mathcal{A}$, there is complete decoupling between layers. This means that the MG convergence speed is determined by the convergence speed on the individual layers. The origin of the "optimal" $\mathcal{A}$-kernel is in works on constructive quantum field theory, see Ref. [25]. For the purpose of numerical computations, it is convenient to determine the optimal $\mathcal{A}$ as solution of the equation

$$([D + \alpha\, C^*C]\mathcal{A})(z,x) = \alpha\, C^*(z,x) \qquad (9)$$

for large $\alpha$. Mack [25] pointed out that it will be essential for beating CSD in interacting models that the layers of an MG decouple as much as possible. This is the case for the ideal $\mathcal{A}$. But, unfortunately, the optimal $\mathcal{A}$ does not fulfill the above mentioned practicality condition, so that the idealized MG algorithm cannot be used for production runs. However, by its use it could be proved numerically [18, 19, 17] that *the MG method works in principle in arbitrarily disordered systems*; see Sec. 4.

## 3   Ground-state Projection Multigrid Method

We assume that the operator $D$ in (1) is connected to a Hamiltonian. A particularly attractive MG method is the "ground-state projection multigrid" approach. In gauge theories (see Sec. 4) this method is covariant.

The central idea of the ground-state projection MG philosophy is that a local action should define the block spin (or $C$, respectively). The averaging operator $C$ from a grid to the next coarser grid is a projector on the ground-state of a block-local Hamiltonian. The idea behind this is that the appropriate notion of smoothness depends on the dynamics, i.e. on $D$, in general. [3]) Results which were found in Ref. [17] confirm the insight that *smooth means little contributions from eigenfunctions to high eigenvalues of D*. This point is important in systems in gauge fields and for other disordered systems.

In order to be concrete, let us choose $D$ as a negative Laplace operator $-\Delta$ in an ordered system. In this case the adjoint $C^*$ of the averaging kernel $C$ satisfies the eigenvalue equation(s),

$$(-\Delta_{b.c.,x} C^*)(z,x) = \lambda_0(x) \, C^*(z,x) \; , \tag{10}$$

together with the subsidiary condition $C^*(z,x) = 0$ if $z \notin x$. $\Delta_{b.c.,x}$ is the lattice Laplacian with "suitable" boundary conditions (b.c.) on the boundary of block $x$ (chosen such that $-\Delta_{b.c.,x}$ is positive semidefinite, see below), and $\lambda_0(x)$ is its lowest eigenvalue. $\Delta_{b.c.,x}$ acts on argument $z$. The solution of the eigenvalue equation is made unique by imposing a normalization condition, either in the form $CC^* = 1\!\!1$ [17] or $< C\, , \, C > = 1\!\!1$ [20]. When one deals with a gauge theory, the normalization condition does not fix the solution of the eigenvalue equation uniquely. In this case one has to impose an additional "covariance condition" [21].

In numerical simulations of lattice field theories [15, 9] one usually works with a lattice with periodic b.c. [4]) In this case we define $\Delta_{b.c.,x}$ with "Neumann b.c." as follows

$$(\Delta_{N,x}\phi)(z) = \sum_{\substack{z' \text{ n.n. } z \\ z' \in x}} [\phi(z') - \phi(z)] \quad \text{for } z \in x. \tag{11}$$

Summation on the r.h.s. of Eq. (11) is over next neighbors $z'$ of $z$ which lie in the same block $x$. The lowest eigenvalue $\lambda_0(x)$ of (11) equals zero for all blocks $x$. Solutions of the eigenvalue equation (10) are constants on blocks. These constants are determined by the normalization condition $CC^* = 1\!\!1$. MG algorithms with piecewise constant restriction and interpolation operators are successfull in eliminating CSD completely in case of free fields [11, 17].

# 4 Multigrid Approaches to the Propagator Problem in Elementary Particle Physics

Monte Carlo simulations of Quantum Chromodynamics (QCD) are an important numerical tool to study the theory of the strong interaction by nonperturbative methods [15, 26]. Presently the only practical and exact algorithm for theories involving dynamical fermions is the "Hybrid

---

[3]) However, the definition of a block-local Hamiltonian requires the specification of boundary conditions (b.c) on block boundaries. The necessity of specifying b.c. is responsible for the fact that "ground-state projection MG" is not an a priori defined scheme. One has to be careful in the definition of b.c. on block boundaries.

[4]) If fermions are involved it is common to use antiperiodic b.c. for Fermi fields in the "time"-direction.

Monte Carlo" algorithm [8]. Massive amounts of supercomputer time are required, and more than 95% of the CPU time is spent for the computation of fermionic propagators in background gauge fields. This computation suffers from CSD as one approaches the continuum limit, and one hopes to overcome CSD by an MG approach.

## 4.1 The problem and ist symmetries

In a numerical simulation of a lattice gauge theory one has to consider the joint collection of gauge and Fermi fields which are stochastically distributed with a certain Boltzmann factor [15, 26]. A (compact) gauge field $U$ is a $U(1)$ or $SU(N_c)$ valued field [5] which is defined on the links $(z, z')$ of the lattice. ($N_c$ is the number of "colors", $z$ and $z'$ are nearest neighbors.) $U(z, z')$ serves as a parallel transporter from $z'$ to $z$. The oppositely orientated link $(z', z)$ carries the gauge field $U(z', z) = U(z, z')^\dagger$. The discretization of Fermi fields is notoriously difficult. Two kinds of lattice fermions are in use nowadays [26]: "Wilson fermions" and "staggered fermions". We will use staggered fermions for illustration. In this case the Fermi field $\phi$, defined on the lattice sites, is an $N_c$-component complex vector: $(\phi^r)_{r=1,\ldots,N_c}$. In $d$ space-time dimensions the covariant staggered Dirac operator $\slashed{D}$ is defined by

$$(\slashed{D}\phi)^r(z) = \frac{1}{a} \sum_{\mu=1}^{d} \eta_\mu(z) \left[ U(z, z + \tfrac{1}{2}e_\mu)^{rs} \phi^s(z + \tfrac{1}{2}e_\mu) - U(z, z - \tfrac{1}{2}e_\mu)^{rs} \phi^s(z - \tfrac{1}{2}e_\mu) \right] \; , \qquad (12)$$

where an implicit summation over the color index $s = 1, \ldots, N_c$ is understood. For notational simplicity we will suppress the color indices in the following. $\eta_\mu$ are the lattice remnants of the Dirac $\gamma$-matrices. They are complex numbers of modulus 1, and may be chosen as $\eta_1(z) = 1$, $\eta_2(z) = (-1)^{n_1}$, $\eta_3(z) = (-1)^{n_1+n_2}$, $\eta_4(z) = (-1)^{n_1+n_2+n_3}$, for $z = \frac{a}{2}(n_1, n_2, n_3, n_4)$. Free staggered fermions enjoy discrete translation invariance under shifts by twice the separation of neighboring sites [16, 22]. Therefore we denote the lattice spacing by $a/2$ in this case. $e_\mu$ denotes a lattice vector of length $a$ in $\mu$-direction.

$\slashed{D}$ is antihermitean so that $-\slashed{D}^2$ is hermitean and positive (semi)definite. Let us assume that we are given a gauge field configuration $U$ as the result of a stochastic process. Then the equation which one has to solve very frequently in a Hybrid Monte Carlo simulation reads

$$\left(-\slashed{D}^2 + m^2\right)\phi = f \qquad (13)$$

where $m$ is a bare mass parameter, and $f$ is given. We invite the reader to work out the explicit form of $\slashed{D}^2$, and to convince oneself that $\slashed{D}$ of Eq. (12) is a square root of a lattice Laplacian in the free case (i.e. for $U = \mathbb{1}$).

An important notion in gauge theories is that of *gauge covariance*. A (local) gauge transformation $g$ is specified on a lattice by a map $g : \Lambda \to G, z \mapsto g(z)$, where $G$ denotes the unitary gauge group. Under a gauge transformation $g$ a matter field $\phi$ transforms according to

$$\phi(z) \mapsto \phi'(z) = g(z)\,\phi(z) \; . \qquad (14)$$

---
[5] The gauge group for Quantum Electrodynamics (QED) is $U(1)$, for QCD it is $SU(3)$.

The transformation law for a link variable $U(z, z')$ is

$$U(z, z') \mapsto U'(z, z') = g(z) U(z, z') g(z')^{-1} . \qquad (15)$$

*The discretized partial differential equation (13) exhibits gauge covariance*, i.e. if $\phi$ is the solution of (13) for given $\{U, f\}$, then $g\phi$ is the solution for the gauge-transformed configuration $\{U', f'\}$.

## 4.2 How to proceed with MG?

In order to apply MG techniques to the propagator problem (or more generally to disordered systems) one has to answer questions like

(i) How to generalize MG methods to gauge theories? The algorithm shall preserve gauge covariance.

(ii) What does smoothness mean in disordered systems?

(iii) How to choose block lattices/coarse grids? This is a priori not clear particularly for staggered fermions.

(iv) etc.

(i) It is possible to preserve gauge covariance in MG algorithms for the solution of Eq. (13). In order to achieve this, the kernels of the restriction and interpolation operator $C(x, z)$ and $\mathcal{A}(z, x)$ have to depend on the gauge field, and they have to become $N_c \times N_c$ matrices [3, 21, 17]. These matrices are not elements of the gauge group, in general.

(ii) The meaning of smoothness in disordered systems is discussed for instance in Refs. [1, 17, 2]. A function $\xi$ on a lattice $\Lambda$ is smooth on length scale $a$ when

$$\|D\xi\| \ll \|\xi\| \qquad (16)$$

in units $a = 1$. This defintion implies that the smoothest function is the lowest eigenmode of $D$. In ordered systems (16) is consistent with a geometrical meaning of smoothness.

(iii) In case of staggered fermions one should use a blocking procedure which is consistent with the lattice symmetries of free fermions [22]. This forces us to choose a blocking factor of $L_b = 3$ (or any other odd number). Even $L_b$ are not allowed. This is remarkable because usually one takes fullest advantage of the MG approach by using a blocking factor of 2.

One may doubt whether the "geometric MG" with prescribed block lattices is an appropriate starting point for problems in disordered systems. One might consider to employ an "algebraic MG" (AMG) approach [28]. However, up to now no generalization of AMG has been found for lattice gauge theories, and the results of Refs. [18, 19] show that *the geometric MG method works in principle in arbitrarily disordered gauge fields*.

| group | operator to be inverted | gauge field | lattice sizes |
|---|---|---|---|
| "Israel" [3, 13, and references therein] 1989–ongoing | $\not{D} + m$ staggered fermions | 2-$d$ $U(1)$ 2-$d$ $SU(2)$ 2-$d$ $SU(3)$ | $\leq 256^2$ $\leq 256^2$ $\leq 128^2$ |
| "Amsterdam" [14, and references therein] 1990–1992 | $-\not{D}^2 + m^2$ staggered fermions staggered fermions and Wilson fermions | 2-$d$ $SU(2)$ 2-$d$ $SU(2)$ | $\leq 128^2$ $\leq 128^2$ |
| "Boston" [7, and references therein] 1990–1991 | $-\Delta + m^2$ | 2-$d$ $U(1)$ 4-$d$ $U(1)$ 2-$d$ $SU(2)$ | $\leq 64^2$ $\leq 16^4$ $\leq 32^2$ |
|  | $(\gamma_\mu + 1)D_\mu + m$ Wilson fermions | 2-$d$ $U(1)$ | $64^2$ |
| [29] 1990–1992 | $(\gamma_\mu + 1)D_\mu + m$ Wilson fermions | 2-$d$ $U(1)$ 4-$d$ $SU(3)$ | $64^2$ $16^4$ |
| "Hamburg" [21, 18, 22, 23, 1, 17, 19, 20, 2, 24] 1990–ongoing | $-\Delta + m^2$ $-\not{D}^2 + m^2$ staggered fermions | 2-$d$ $SU(2)$ 4-$d$ $SU(2)$ 2-$d$ $SU(2)$ 4-$d$ $SU(2)$ | $\leq 128^2$ $\leq 18^4$ $\leq 162^2$ $\leq 18^4$ |

Table 1: *Overview of works on MG methods for propagators in lattice gauge theories.*

## 4.3 Overview of existing works

Big efforts have been undertaken to find efficient MG methods for the computation of propagators in background gauge fields. We give an overview of these works in Table 1. All works mentioned in Table 1, except Ref. [24], focused only on quenched gauge fields. This means that the quarks are treated as static in the Monte Carlo updating procedure.

### 4.3.1 Works of the Israel group

The Israel group introduced an MG method which they called parallel-transported multigrid ("PTMG") [3, and references therein]. This method implements gauge covariance directly by defining integral kernels through weighted averages of parallel transporters on finer grids. Blocking was done with a factor of $L_b = 2$, which is not consistent with the symmetries of free staggered fermions, but which is legitimate from a purely computational point of view. Conclusions in Refs. [3, 13] are that PTMG outperforms the commonly used algorithms very close to the continuum limit on 2-$d$ lattices with gauge groups $U(1)$, $SU(2)$ and $SU(3)$.

### 4.3.2 Works of the Amsterdam group

The Amsterdam group [14, and references therein] used a covariant ground-state projecting MG method in 2-$d$ $U(1)$, which they gave up in 2-$d$ $SU(2)$ where they used an algorithm with gauge fixing. Blocking was done with a factor of $L_b = 2$ in case of staggered fermions (but differently from [3]); Wilson fermions were blocked with a factor of 4 in the first blocking step, then with a factor of 2. Conclusions are that in the 2-$d$ systems investigated MG is comparable with the CG algorithm only for very large correlation lengths.

### 4.3.3 Works of the Boston group

The Boston group [7, and references therein] used a covariant ground-state projecting MG method for 2-$d$ $U(1)$ bosonic propagators. They have a "variational-state projection" which is not ground-state projecting but covariant for bosonic propagators in 2-$d$ $SU(2)$ and 4-$d$ $U(1)$, and for Wilson fermions in 2-$d$ $U(1)$. No competitive MG algorithm was found.

V. Vyas [29] made his Ph.D. in Boston but his work is independent of the authors of Ref. [7]. Vyas's method is a PTMG approach where the weights for the different paths, over which parallel transporters are averaged, are fixed by a modified Migdal-Kadanoff renormalization group transformation. Vyas concludes a reduction of CSD for large correlation lengths in 2-$d$ $U(1)$. He claims that his method is competetive for current lattice sizes in QCD (4-$d$ $SU(3)$), but the present author is unaware whether this has been confirmed later on.

### 4.3.4 Works of the Hamburg group

The Hamburg group used exclusively gauge covariant ground-state projecting MG schemes. The present author showed that the method is numerically implementable in four-dimensional nonabelian gauge fields [21]; no gauge fixing is required. In case of the bosonic problem one wants to invert $-\Delta + m^2$ where $\Delta$ is the *gauge covariant lattice Laplacian* defined through

$$(\Delta\phi)(z) = \sum_{z' \text{ n.n. } z} [U(z,z')\phi(z') - \phi(z)] \; . \tag{17}$$

The (adjoint of the) averaging kernel $C$ fulfills the covariant eigenvalue equation(s)

$$(-\Delta_{N,x} C^*)(z,x) = \sum_{\substack{z' \text{ n.n. } z \\ z' \in x}} [C^*(z,x) - U(z,z')C^*(z',x)] = \lambda_0(x) C^*(z,x) \tag{18}$$

for $z \in x$. Remember that (18) is an equation for $N_c \times N_c$ matrices. In case of staggered fermions one can define $C$ through similar eigenvalue equations but in order to avoid too much technicalities here we refer the reader to Refs. [22, 17, 20]. A comprehensive summary about the computation of propagators by means of various algorithms can be found in [17]. In case of bosonic propagators the CG algorithm can be outperformed in 4-$d$ $SU(2)$ gauge fields on lattices $\gtrsim 18^4$ [18]. MG methods work for staggered fermions in 4-$d$ $SU(2)$ gauge fields, but for realistic lattice sizes simple MG methods are inferior to CG [23, 17]. However, by means of the idealized MG algorithm it was shown that *in principle MG methods are able to eliminate CSD in 4-d nonabelian gauge fields, both for bosonic propagators [18] and for staggered fermions [19]*.

A recent proposal by Bäker [2] is the "iteratively smoothing unigrid (ISU)". This practical algorithm takes care of the fact that low-lying modes in disordered systems are localized, and it is successful in eliminating CSD for bosonic propagators in 2-d $SU(2)$ gauge fields. Work for fermions is in progress.

### 4.4 Finally some words about preconditioning and spectral properties

Attempts to precondition the inversion of the staggered fermion matrix were not successful in conventional iterative algorithms [27]. No preconditioning was employed in any of the above mentioned works on MG methods for propagators. In Ref. [7] the authors announced to test their MG method as a preconditioner for CG but nothing has been published on this issue.

In case of free fields ($U \equiv 1$) MG is able to eliminate CSD completely. Hence, no preconditioning is necessary. The eigenvalues of the staggered $(-\slashed{D}^2 + m^2)$ are clustered in this case. In nontrivial gauge fields the eigenvalues are distributed uniformly between the smallest and the largest eigenvalue, see Fig. 1. $(-\slashed{D}^2 + m^2)$ has condition numbers $\kappa$ of order $O(10^2)$–$O(10^4)$ in an interesting mass range [24]. The (asymptotic) convergence behavior of the CG algorithm depends only on $\kappa$ and on the lattice size. It is demanding to devise preconditioned CG algorithms with much smaller $\kappa$'s. The behavior of MG algorithms is not affected by $\kappa$ but depends on the spectrum in a more subtle way. A lot of work remains to be done for an inversion of $(-\slashed{D}^2 + m^2)$ without CSD. The problem is that there are (very) many approximate zero modes. In addition, these low-lying modes are localized [20, 2]. Perhaps a recent general proposal of recombining MG iterants can help [5].

ACKNOWLEDGMENTS

It is a pleasure to thank the organizers of "Modelling 94" for their kind invitation and for providing the framework for a stimulating atmosphere at the conference. Financial support of my work by Deutsche Forschungsgemeinschaft is gratefully acknowledged. (Present grant Wo 389/3-1.)

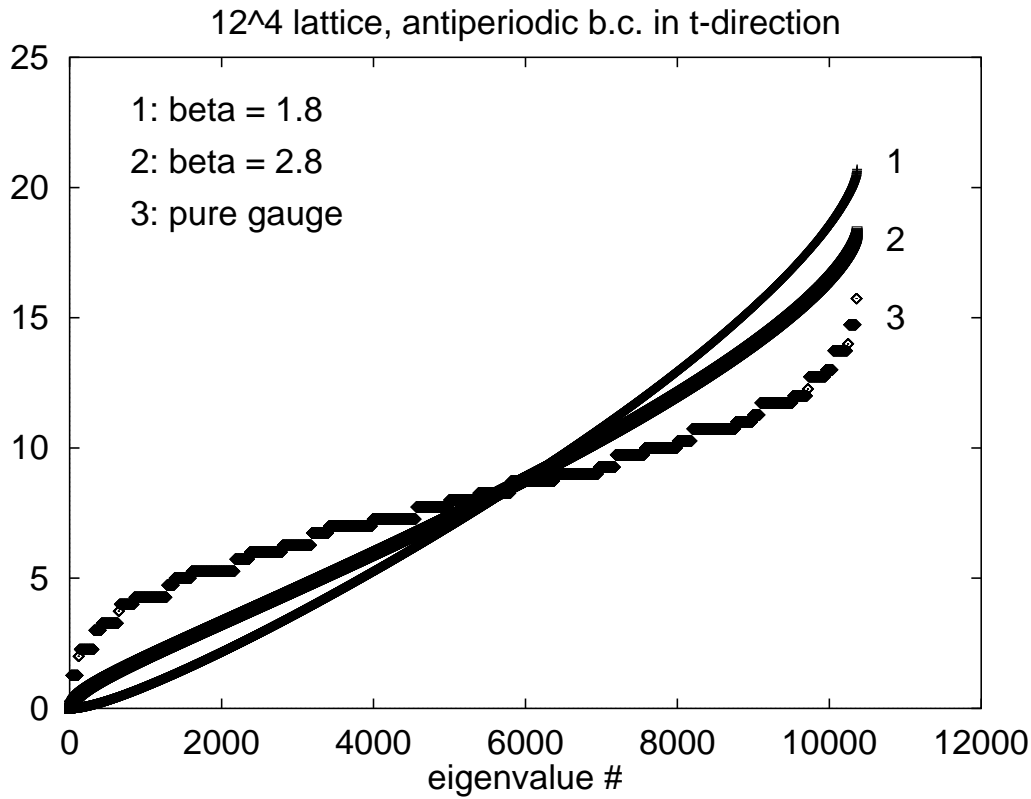

Figure 1: *Spectrum of $-\slashed{D}^2$ in $SU(2)$ gauge fields on $12^4$ lattices with antiperiodic boundary conditions. "beta" is the coupling constant in the Wilson action, cf. [15], "pure gauge" is the trivial case $U \equiv \mathbb{1}$.*